\documentstyle[12pt,epsfig]{article}
\newlength{\dinwidth}
\newlength{\dinmargin}
\newcommand{\ba}{\begin{array}}
\newcommand{\ea}{\end{array}}
\newcommand{\be}{\begin{eqnarray}}
\newcommand{\ee}{\end{eqnarray}}

\newcommand{\sq}{\sqrt{2}}
\newcommand{\half}{\frac{1}{2}}
\newcommand{\del}{\partial}
\newcommand{\ra}{\rightarrow}
\newcommand{\lt}{\widetilde\lambda}
\newcommand{\ol}{\overline}
\def\Z{{\bf Z}}
\def\cT{{\cal T}}
\def\cF{{\cal F}}
\def\cD{{\cal D}}
\def\cQ{{\cal Q}}
\def\cL{{\cal L}}

\def\vectt[#1,#2]{\left(%
\begin{array}{c} #1 \\ #2 \end{array} \right)}

\def\trivectt[#1,#2,#3]{\left(%
\begin{array}{c} #1 \\ #2 \\ #3 \end{array} \right)}

\newcommand{\VEV}[1]{\left\langle #1\right\rangle}

\newcommand{\ket}[1]{\left|\, #1\,\right\rangle}

\def\nn{\nonumber \\}
\def\l{\left}
\def\r{\right}
\def\d{{\rm d}}
\def\wt{\widetilde}
\def\wh{\widehat}
\def\wht{\widehat{x}^0}
\def\whx{\widehat{x}^1}

\newcommand{\gsim}{\mathrel{\mathop{\kern 0pt \rlap
  {\raise.2ex\hbox{$>$}}}
  \lower.9ex\hbox{\kern-.190em $\sim$}}}
\newcommand{\lsim}{\mathrel{\mathop{\kern 0pt \rlap
  {\raise.2ex\hbox{$<$}}}
  \lower.9ex\hbox{\kern-.190em $\sim$}}}

\begin{document}
\thispagestyle{empty} \addtocounter{page}{-1}
\begin{flushright}
SNUST 030301\\
{\tt hep-th/0303133}\\
\end{flushright} \vspace*{1cm}
\centerline{\Large \bf  Rolling of Modulated Tachyon with Gauge
Flux} \vskip0.4cm \centerline{\Large \bf and} \vskip0.4cm
\centerline{\Large \bf Emergent Fundamental String ~\footnote{Work
supported in part by the BK-21 Initiative in Physics (SNU
Project-2), the KRF Overseas Research Grant, the KOSEF
Interdisciplinary Research Grant 98-07-02-07-01-5, and the KOSEF
Leading Scientist Grant.}} \vspace*{1.5cm}
\centerline{\bf Soo-Jong Rey${}^{1,2}$ {\rm and} Shigeki
Sugimoto${}^3$} \vspace*{1.0cm}
\vskip0.4cm
\centerline{\it School of Natural Sciences, Institute for Advanced
Study} \vspace*{0.2cm}
\centerline{\it Einstein Drive, Princeton NJ 08540 \rm USA ${}^1$}
\vskip0.4cm
\centerline{\it School of Physics \& Center for Theoretical
Physics} \vspace*{0.2cm}
\centerline{\it Seoul National University, Seoul 151-747 \rm
KOREA${}^2$}
\vskip0.4cm
 \centerline{\it The Niels Bohr Institute}
\vspace*{0.2cm}
\centerline{\it Blegdamsvej 17, DK-2100 Copenhagen \rm DENMARK
${}^3$} \vspace*{0.8cm}
\centerline{\tt sjrey@gravity.snu.ac.kr, sugimoto@nbi.dk}
\vskip1cm \centerline{\bf abstract} \vspace*{0.3cm}
We investigate real-time tachyon dynamics of unstable D-brane
carrying fundamental string charge. We construct the boundary
state relevant for rolling of modulated tachyon with gauge fields
excited on the world-volume, and study spatial distribution of the
fundamental string charge and current as the D-brane decays. We
find that, in contrast to homogeneous tachyon rolling, spatial
modulation of the tachyon field triggers density wave of strings
when electric field is turned on, and of string anti-string pairs
when magnetic field is turned on. We show that the energy density
and the fundamental string charge density are locked together, and
evolve into a localized delta-function array (instead of evolving
into a string fluid) until a critical time set by rolling
tachyon's initial condition. When the gauge fields approach the
critical limit, the fundamental strings produced become BPS-like.
We also study the dynamics via effective field theory, and find
agreement.


\baselineskip=18pt
\newpage

\section{Introduction}
Recently, there has been considerable progress in understanding
decay of unstable D-brane in string theory \footnote{Among various
works contributed to the progress, a list of those more relevant
to the present work includes~\cite{sen}--\cite{IsUe2}.}. The decay
proceeds, as analyzed first by Sen \cite{sen}, via rolling of the
tachyon living on the D-brane world-volume from the top of the
tachyon potential to the bottom (closed string vacuum). It was
found that, in the weak coupling limit $g_{\rm st} = 0$, as the
tachyon rolls down the potential hill, the unstable D-brane is
converted to a pressureless gas, referred as ``tachyon matter".
Details of the decay process is of considerable physical interest,
as it would shed light on semi-classical and non-perturbative
aspects of string {\sl dynamics}.

An interesting physics question is, as the D-brane decays, how
various constituents bound inside an unstable D-brane are
liberated. To answer the question, we will consider a bound-state
of unstable D-brane with fundamental string, and study dynamical
evolution of energy and charge distributions. In the previous
works \cite{MuSen,reysugimoto}, this problem was studied for
homogeneous rolling of the tachyon, and it was found that the
constituents form a uniformly distributed fundamental string fluid
on the hyper-surface of the D-brane world-volume, behaving
BPS-like in the limit we take the world-volume gauge fields to a
critical value. In this work, we shall examine whether and, if it
does, how {\sl spatial modulation} of the rolling tachyon field
would affect fate of constituent's distribution in the ambient
space-time. In approaching the problem, we shall continue
utilizing the approach set forth in the previous work
\cite{reysugimoto}.

We begin with salient features concerning homogeneous rolling of
the open string tachyon. The real-time dynamics is described
\cite{sen}, in the boundary state approach, by turning on an
appropriate boundary interaction to the $c=1$ conformal field
theory of the $X^0$-field:
\be \Delta S = \widetilde{\lambda} \int \d t \oint \d \sigma \,
\delta(t) \cosh X^0(t, \sigma). \nonumber \ee
Boundary state corresponding to the above interaction was
constructed in \cite{sen}, and takes the form
\be \vert {\rm D25} \rangle_{T(x^0)} = \vert B \rangle_{X^0}
\otimes_{i=1}^{25} \vert N \rangle_{X^i} \otimes \vert {\rm ghost}
\rangle, \label{c1bs} \ee
where
\be \vert B \rangle_{X^0}
&=& f (\widehat{x}^0) \vert 0 \rangle +
g(\widehat{x}^0) \alpha^0_{-1} \overline{\alpha}^0_{-1} \vert 0
\rangle \nn
&+& \left[ h_1 (\widehat{x}^0) \alpha^0_{-2}
\overline{\alpha}^0_{-2} + h_2(\widehat{x}^0) (\alpha^0_{-1})^2
(\overline{\alpha}^0_{-1})^2 + h_3(\widehat{x}^0) \left(
(\alpha^0_{-1})^2 \overline{\alpha}^0_{-2} + \alpha^0_{-2}
(\overline{\alpha}^0_{-1})^2 \right) \right] \vert 0 \rangle \nn
&+& \cdots \label{b0}\\
\vert N \rangle_{X^i} &=& \exp \left( -
\sum_{n=1}^\infty {1 \over n} \alpha^i_{-n}
\overline{\alpha}^i_{-n} \right) \vert 0 \rangle \quad \qquad
(i=1, \cdots, 25) \nn
\vert {\rm ghost} \rangle &=& \exp \left( - \sum_{n=1}^\infty
\left(\overline{b}_{-n} c_{-n} + b_{-n} \overline{c}_{-n}\right)
\right) (c_0 + \overline{c}_0) c_1 \overline{c}_1 \vert 0 \rangle
\ . \nonumber \ee
The coefficient functions $f(x^0), g(x^0), h_1(x^0), h_2(x^0),
h_3(x^0), \cdots$ describe time evolution of the rolling tachyon
in the closed string channel, and are computable explicitly, as
was done in \cite{sen,MuSen,okudasugimoto}. Expanding the
boundary-state Eq.(\ref{c1bs}) in oscillator levels, one can
obtain explicit form of the linear coupling of the tachyon-rolling
D25-brane to each closed string mode. It was found that, at late
time, the rolling tachyon evolves into a pressureless ``tachyon
matter". It was further found that the coupling to higher-level
closed string states is hierarchically pronounced
\cite{okudasugimoto}, a feature escalated even further in a
situation the world-volume electric and magnetic fields are turned
on \cite{MuSen,reysugimoto}. It also entailed an issue on the
stability of the ``tachyon matter'' once a non-zero $g_{\rm st}$
is turned on.

The rolling of spatially modulated tachyon field
can also be studied using the boundary state approach \cite{seninhomo}.
In \cite{seninhomo}, Sen considered
the inhomogeneous tachyon corresponding to
the following boundary interaction:
\be \Delta S = \widetilde{\lambda} \oint \d \sigma \cosh {X^0
\over \sqrt{2}} \cos {X^1 \over \sqrt{2}}.\nonumber
\ee
It was found that half of the total energy start to accumulate at
localized points and form an array of delta-function-like
singularity at finite time. Furthermore, after the system hits the
singularity, the total energy suddenly reduces to zero and the
energy density vanishes eventually everywhere.

In this paper, we turn on gauge fields along with the
inhomogeneous rolling tachyon, and study time evolution of the
decaying D-brane. Surprisingly, we find that, in the presence of
spatial modulation of the tachyon field, magnetic field (as well
as electric field) induces fundamental string constituents on the
D-brane world-volume. We also find that time evolution develops a
singularity not only for the energy density but also for the
fundamental string charge density. Consequently, our study
indicates the following picture of the D-brane decay. Initially,
driven by the tachyon field modulation, the fundamental string
energy and charge densities evolve into free streaming of thin
flux-tubes, whose thickness approaches zero up to a critical time
set by the initial condition of the tachyon and gauge fields. The
critical time is the moment a singularity develops to the boundary
state --- in particular, energy and charge densities are
discontinuous across the critical time, leading allegedly to
energy and charge non-conservation. Physically, as the energy and
the string charge ought to be conserved, a possibility is that
{\sl macroscopic} fundamental strings of zero thickness are built
up until the critical time and then are decoupled from the system
right after the critical time (recall that we are working with
$g_{\rm st}$ set to zero.).

We study couplings to the massless as well as some of the
lower-level massive closed string modes, which can be extracted
from the boundary state. Through these couplings, the boundary
state acts as a space-time varying source for the closed string
field in the linearized equation of motion
\begin{eqnarray}
(Q_B+\ol Q_B)\ket{\Phi}=\ket{B},
\label{sfteq}
\end{eqnarray}
where $Q_B$ is the BRST operator and $\ket{\Phi}$ is the closed
string field. We find that the coupling to the closed strings
generally blows up sharply as the system evolves across the
critical time, suggesting that the source $\ket{B}$ of the closed
string field equation in Eq.(\ref{sfteq}) may be approximated as a
sort of {\sl impulse}. Then, once $g_{\rm st}$ is turned on, the
equations of motion for both open and closed strings will be
altered by nonlinear interactions, and our analysis may cease to
hold close to the critical time. The situation is analogous to the
formation of a heavy star in general relativity: once the energy
density becomes huge compared to the Planck scale, one cannot
ignore the nonlinear terms in the Einstein field equation.

Several previous works \cite{yietal2,yietal3,kawaikuroki} examined
fate of the fundamental string constituents using the effective
field theory approximation to an unstable D-brane, and claimed
that a diffusive string fluid of arbitrary density profile is
formed at the tachyon potential minimum, at least at the classical
level. Our boundary state (as well as effective field theory)
analysis shows the contrary that thin string-like flux tubes can
be formed via real-time tachyon rolling \footnote{This might be
related to the result of \cite{hoetal}.}.

This paper is organized as follows. In section 2, we begin with
recapitulating Sen's construction of the boundary state
representing rolling of modulated tachyon in bosonic string
theory, and examine couplings to massless and massive closed
string modes. We present detailed analysis of the time evolution
of these couplings, and indicate that the couplings to both
massless and massive closed string modes exhibit a singularity at
the critical time. In section 3, adopting the recipe prescribed in
\cite{reysugimoto}, we construct the boundary state describing
rolling of modulated tachyon with constant electric and magnetic
fields. We extract time evolution of energy and charge density
associated with fundamental string and analyze the behavior. In
section 4, we study the tachyon rolling dynamics again, but now in
the effective field theory approach. We find similar but somewhat
differing result compared to those obtained from the boundary
state approach. We contrast our results to earlier works
\cite{yietal2,kawaikuroki}, and assert that stability criterion
used in these works needs to be reconsidered in our case.
Section 5 is devoted
to discussion including some speculation concerning the physics
around the singularity.
\section{Rolling of Modulated Tachyon}
\subsection{Boundary State Construction}
Following \cite{seninhomo}, consider the boundary-state
description of the rolling of inhomogeneous tachyon on a D25-brane
in the bosonic string theory. In the boundary state approach, it
amounts to turning on the following boundary action:
\be \Delta S = \widetilde{\lambda} \oint \d \sigma \cosh {X^0
\over \sqrt{2}} \cos {X^1 \over \sqrt{2}} \label{c2bdryint} \ee
to the $c=2$ conformal field theory (CFT) associated with $X^0,
X^1$ of the bosonic D25-brane \cite{solvablecft1,solvablecft2}.
The corresponding boundary state describes a D25-brane whose
world-volume tachyon field is modulated sinusoidally in space
$x^1$, and rolling in time $x^0$. Schematically, it takes the
form:
\be \vert {\rm D25} \rangle_{T(x^0,x^1)} &=& \vert B \rangle_{X^0,
X^1} \otimes_{i=2}^{25} \vert N \rangle_{X_i} \otimes \vert {\rm
ghost} \rangle. \label{bs} \ee
The boundary action Eq.(\ref{c2bdryint}) turns out to define an
exactly solvable perturbation, as the $c=2$ CFT boundary state
$\vert B \rangle_{X^0, X^1}$ is obtainable in a factorized form:
\be \vert B \rangle_{X^0, X^1} = \vert B \rangle_+ \otimes \vert B
\rangle_- \ , \label{factorized} \ee
where $\vert B \rangle_\pm$ are the $c=1$ boundary state
Eq.(\ref{b0}) with the replacement of $X^0 \rightarrow (X^0 \pm i
X^1)/\sqrt{2}$ and $\wt{\lambda} \rightarrow \wt{\lambda}/2$
\cite{seninhomo}. Explicitly, we find that
\be \vert B \rangle_\pm &=& f^\pm(\wht, \whx) \vert 0 \rangle \nn
&+& {1 \over 2} g^\pm(\wht, \whx) ( \alpha^0_{-1} \pm i
\alpha^1_{-1} )( \overline{\alpha}^0_{-1} \pm i
\overline{\alpha}^1_{-1}) \vert 0 \rangle \nn
&+& {1 \over 4} h_1^\pm (\wht, \whx) (\alpha^0_{-2} \pm i
\alpha^1_{-2} ) (\overline{\alpha}^0_{-2} \pm i
\overline{\alpha}^1_{-2} ) \vert 0 \rangle \nn
&+& {1 \over 8} h_2^\pm(\wht, \whx) (\alpha^0_{-1} \pm i
\alpha^1_{-1})^2 ( \overline{\alpha}^0_{-1} \pm i
\overline{\alpha}^1_{-1} )^2 \vert 0 \rangle \nn
&+& {i \over 4} h_3^\pm (\wht, \whx) \left[ (\alpha^0_{-1} \pm i
\alpha^1_{-1})^2 (\overline{\alpha}^0_{-2} \pm i
\overline{\alpha}^1_{-2} ) + (\alpha^0_{-2} \pm i \alpha^1_{-2})
(\overline{\alpha}^0_{-1} \pm i \overline{\alpha}^1_{-1})^2
\right] \vert 0 \rangle \nn &+& \cdots  \nonumber \ee
  where the coefficient functions are defined by
\be f^\pm (x^0, x^1) &=& \left(1 + e^{+{x^0 \pm i x^1\over
\sqrt{2}}} \sin (\widetilde{\lambda} \pi /2) \right)^{-1} + \left(
1 + e^{-{x^0 \pm i x^1\over \sqrt{2} }} \sin (\widetilde{\lambda}
\pi/2) \right)^{-1} - 1, \label{fpm} \\
g^\pm (x^0, x^1) &=& 2\cos^2 (\widetilde{\lambda} \pi/2 ) - f^\pm
(x^0, x^1), \label{gfromf} \\
h^\pm_1(x^0, x^1) &=& 2\cos^2 (\wt{\lambda} \pi/2) - 2 \sin
(\wt{\lambda} \pi/2) \cos^2 (\wt{\lambda} \pi/2) \cosh
\left({x^0\pm i x^1 \over \sqrt{2}}\right) -  f^\pm(x^0, x^1), \nn
h^\pm_2(x^0, x^1) &=& 4 \sin(\wt{\lambda} \pi/2) \cos^2
(\wt{\lambda} \pi/2) \cosh \left({x^0 \pm i x^1 \over
\sqrt{2}}\right) + f^\pm(x^0, x^1), \nn
h^\pm_3(x^0, x^1) &=& - 2 \sin (\wt{\lambda} \pi/2) \cos^2
(\wt{\lambda} \pi/2) \sinh \left({x^0 \pm i x^1 \over \sqrt{2}}
\right). \label{h3}\ee
Note that the functions labeled with $+$ superscript are complex
conjugate of those labeled with $-$ superscript. Note also that,
except $h^\pm_3(x^0, x^1)$, all coefficient functions are related
to $f^\pm(x^0, x^1)$.

Expanding the matter part of the boundary state Eq.(\ref{bs}) in
powers of the oscillators and coupling functions associated with
them, we obtain
\be \vert \mbox{D25} \rangle_{T} &=& F(x^0, x^1) \vert 0 \rangle
\nn &+& G_{ab}(x^0, x^1) \, \alpha^a_{-1} \overline{\alpha}^b_{-1}
\vert 0 \rangle + {1 \over 2} H_{ab}(x^0, x^1) \, \alpha^a_{-2}
\overline{\alpha}^b_{-2} \vert 0 \rangle \nn &+& {1 \over 4}
I_{abcd}(x^0, x^1) \, \alpha^a_{-1} \alpha^b_{-1}
\overline{\alpha}^c_{-1} \overline{\alpha}^d_{-1} \vert 0 \rangle
+ {i \over 2} J_{abc}(x^0, x^1) \left(\alpha_{-1}^a \alpha^b_{-1}
\overline{\alpha}^c_{-2} + \alpha_{-2}^c \overline{\alpha}_{-1}^a
\overline{\alpha}_{-1}^b \right) \vert 0 \rangle \nn &+& \cdots,
\label{expandedbs} \ee
where the coupling functions are given by
\be
&&F = \vert\vert f^+ \vert\vert^2, \nn
&&G_{00} = - G_{11} = {\rm Re} (f^+ g^-),~~~ G_{01} = +G_{10}= {\rm Im} (f^+ g^-),\nn
&&G_{ij} = - \vert\vert f^+\vert\vert^2 \delta_{ij} \qquad ({\rm
for}~~i,j = 2, 3, \cdots, 25), \label{FG} \ee
for the tachyon and massless level modes, and
\be &&H_{00} = - H_{11} = {\rm Re} (f^+ h_1^-),~~~ H_{10} =
+H_{01} = {\rm Im} (f^+ h_1^- ),\nn
&&H_{ij} = - \vert\vert f^+\vert\vert^2 \delta_{ij} \qquad ({\rm
for}~~ i,j = 2, 3, \cdots, 25), \nn
&&I_{0000} = I_{1111} = +{\rm Re} (f^+h_2^-) + || g^+\vert\vert^2,\nn
&&I_{0011} = I_{1100} = - {\rm Re} (f^+ h_2^-) + \vert\vert
g^+ \vert\vert^2, \nn
&& I_{0101} = +I_{1010} = +I_{0110} = +I_{1001} = - {\rm Re} (f^+
h_2^-),\nn && I_{0100} = - I_{0111} = +I_{0001} = - I_{1101} =
+{\rm Im} (f^+ h_2^-), \nn &&I_{ijkl} = \vert\vert f^+\vert\vert^2
(\delta_{ik} \delta_{jl} + \delta_{il} \delta_{jk} ) \qquad ({\rm
for}~~i,j,k,l =2, 3, \cdots, 25),\nn && J_{001} = - J_{111} = {\rm
Re} (f^+ h_3^-),~~~  J_{000} = - J_{110} = - {\rm Im} (f^+ h_3^-),
\nn &&J_{010} = +J_{100} = {\rm Re} (f^+ h_3^-),~~~ J_{011} =
+J_{101} = +{\rm Im} (f^+ h_3^-), \label{coefficients} \ee
etc. for the first massive level modes. Various relations among
the coupling functions are attributable to the specific factorized
form of the boundary state, as given in Eq.(\ref{factorized}).

\subsection{Time Evolution of Coefficient Functions}
As alluded above, the D25-brane boundary state $\ket{\rm D25}_T$
acts as a source to the closed string field equation
Eq.(\ref{sfteq}). The source being space-time varying, we thus
need to examine its time evolution. In Eq.(\ref{expandedbs}), the
source is decomposed into closed string mass-levels, and we shall
be interested in the evolution of the coupling functions
Eqs.(\ref{FG}, \ref{coefficients}).

As $f^\pm(x^0, x^1)$ are the building blocks for most of these
coefficients, we will analyze their behavior first. From
Eq.(\ref{fpm}), we see that they are singular at $ x^0\pm
ix^1=\mp\sq\log\left(-{\sin(\lt\pi/2)}\right)$. To be specific,
consider the case $0<\sin(\lt\pi/2)<1$ in the following. Flipping
the sign of $\lt$, interpreted as flipping the sign of the initial
value of the tachyon field, amounts to relocating the $x^1$-locus
of the singularities by $\sq\pi$. When $0<\sin(\lt\pi/2)<1$, the
singularities are located at $x^1=\sq(2n+1)\pi$ ($n\in\Z$) at the
critical time $x^0=\pm x_c^0$, defined by \be
x_c^0=\sq\log\left({1 \over |\sin(\lt\pi/2)|}\right).
\label{critical} \ee

Let us examine closely the behavior of $f^\pm(x^0, x^1)$ around
the critical time. Useful formulae for our analysis are:
\be
\lim_{\beta \rightarrow 1 \mp 0} \left( {1 \over 1 +
\beta\, e^{i x/\sq}}+{1 \over 1+\beta\, e^{-ix/\sq}} \right) &=& 1
\pm 2\sq\pi \sum_{n\in\Z} \delta \l(x-\sq(2n+1)\pi\r),
\label{formula0} \ee
and
\be \int_0^{2\sq\pi} \d x^1\, {1 \over 1 + \alpha\, e^{\pm i x^1/\sq}} =
\left\{ \begin{array}{ccc} 0 & {\rm for} & \vert \alpha \vert > 1
\\ 2\sq \pi & {\rm for} & \vert \alpha \vert < 1 \end{array} \right. .
\label{formula1} \ee

The behavior of the real part of $f^\pm(x^0,x^1)$ close to the
critical time $x^0 = x_c^0$ is obtainable by making use of
Eq.(\ref{formula0}):
\be && \lim_{x^0 \rightarrow x_c^0 \mp 0}\left(
 f^+(x^0,x^1)+ f^-(x^0,x^1)\right)
\nn &=&\pm 2\sq \pi
\sum_{n\in\Z}\delta\left(x^1-\sq(2n+1)\pi\right)\nn && + \left[
\left( 1 + \sin^2 (\widetilde{\lambda} \pi /2 )e^{ix^1/\sq}
\right)^{-1} +\left( 1 + \sin^2 (\widetilde{\lambda} \pi /2 )e^{-i
x^1/\sq} \right)^{-1} - 1 \right]. \label{realpart} \ee
Note that the sign of the $\delta$-function part (the first line)
flips as $x^0$ passes through the critical time $x_c^0$,
 whereas the regular part (the second line)
remains the same. Note also that, for a fixed $x^0 \sim x^0_c$,
the real part Eq.(\ref{realpart}) is an even function of $x^1$
across each singularity located at $x^0 = \sqrt{2} (2n+1) \pi$.
The imaginary part of $f^\pm(x^0, x^1)$ is also singular at the
critical time $x^0_c$:
\be && \lim_{x^0 \rightarrow x_c^0}\left( f^+(x^0,x^1)-
f^-(x^0,x^1)\right) \nn &= & - i \tan \left({x^1 \over 2\sq}
\right) \nn &&+ \left[ \left( 1 + \sin^2 (\widetilde{\lambda} \pi
/2 )e^{ix^1/\sq} \right)^{-1} -\left( 1 + \sin^2
(\widetilde{\lambda} \pi /2 )e^{-i x^1/\sq} \right)^{-1} \right] \
. \label{imagpart} \ee
In this case, there is no discontinuity in $x^0$ for both the
singular part (the first line) and the regular part (the second
line). For a fixed $x^0 \sim x^0_c$, the imaginary part
Eq.(\ref{imagpart}) is an odd function of $x^1$ across each
singularity at $x^0 = \sqrt{2} (2n+1) \pi$.

It is also of interest to examine the spatial averages of the
functions $f^\pm(x^0,x^1)$. We obtain them from
Eq.(\ref{formula1}) as
\be \langle f^\pm \rangle (x^0) := \frac{1}{2 \sq \pi}
\int^{2\sq\pi}_0 dx^1 \, f^\pm (x^0,x^1) = \left\{
\begin{array}{ccc} 0 & {\rm for} & \vert x^0 \vert > x_c^0 \\
1 & {\rm for} & \vert x^0 \vert < x_c^0 \end{array} \right. .
\label{ave}
\ee

Generically, functions associated with massive modes, such as
$h_1^\pm, h_2^\pm$, depend on $f^\pm$ (see Eq.(\ref{h3})), and
hence develop the same singularity as $f^\pm$ at the critical time
$x^0_c$. There are, however, some exceptions and some other
functions, such as $h_3^\pm$, do not depend on $f^\pm$, and evolve
in time in a regular manner.

\subsection{Coupling to Closed String Modes}
\label{coupling}
Having obtained the coupling functions Eqs.(\ref{FG},
\ref{coefficients}), let us now examine time evolution of each of
them.

\begin{figure}[tb]
   \vspace{0cm}
   \centerline{
   \epsfysize=6cm  \epsffile{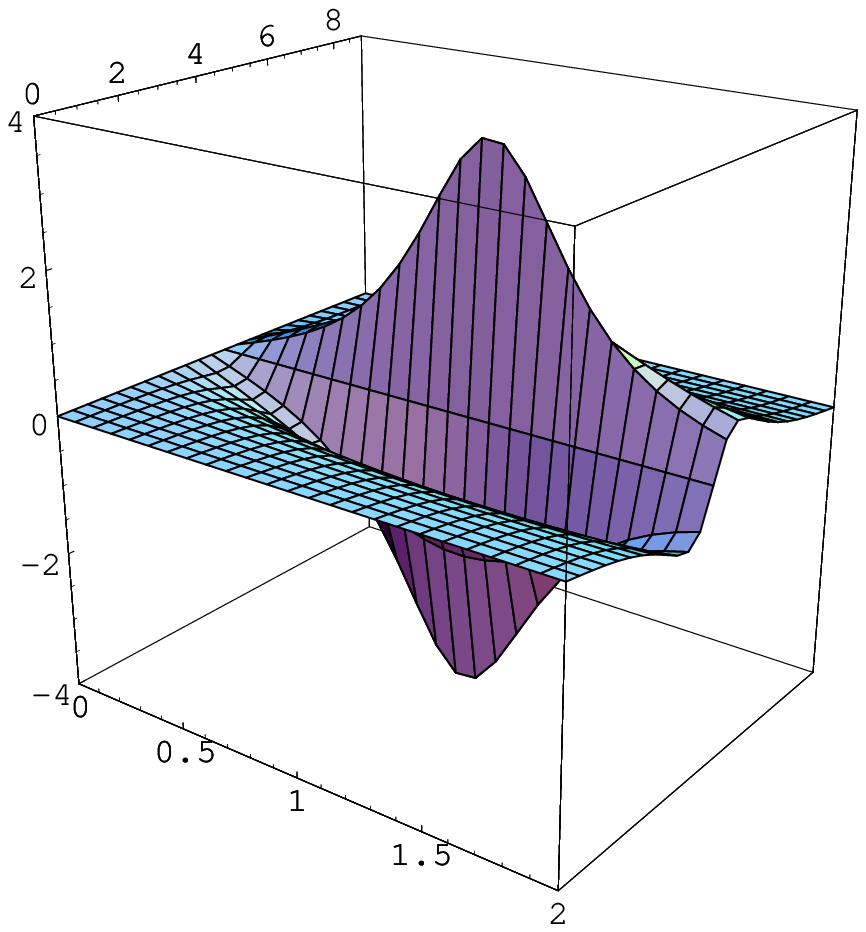}\hskip1.5cm
   \epsfysize=5.5cm  \epsffile{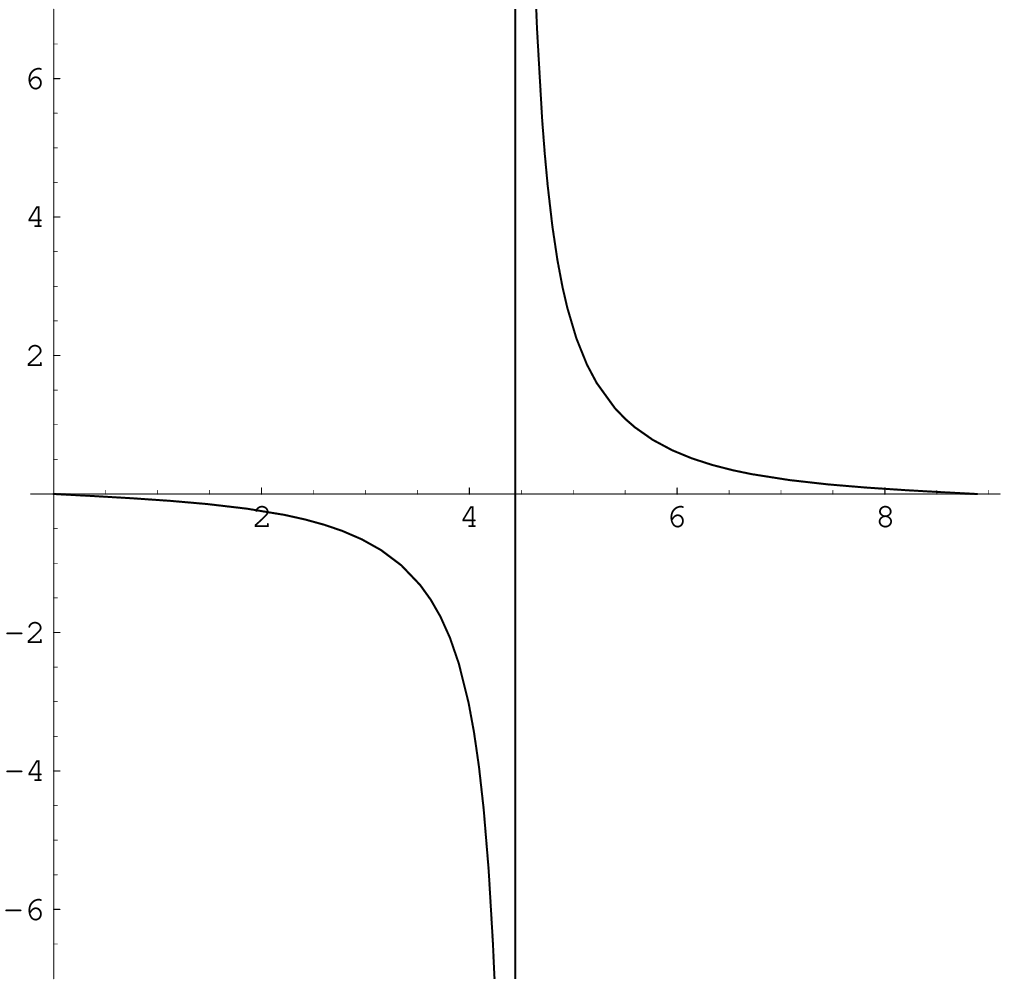}}
\caption{\sl Behavior of the momentum density ($\propto {\rm
Im}f^+(x^0,x^1)$) (vertical axis in arbitrary unit) for $\lt=0.3$
in $x^0 = [0, 2]$ and $x^1 = [0, 2\sq\pi]$ in the left, and its
section at the critical time $x_c^0\sim 1.116\cdots$
 in the right.} \label{momfig}
\end{figure}

$\bullet$ \underline{graviton coupling} \hfill\break
The energy-momentum tensor of the decaying D25-brane can be read
from the boundary state Eq.(\ref{expandedbs}) as \cite{sen}
\be T_{ab}(x^0, x^1) = \frac{{\cal T}_{25}}{2} \left(
G_{ab}(x^0, x^1) - \eta_{ab} F(x^0, x^1) \right).
\nonumber \ee
\begin{figure}[tb]
   \vspace{0cm}
   \centerline{
   \epsfysize=6cm  \epsffile{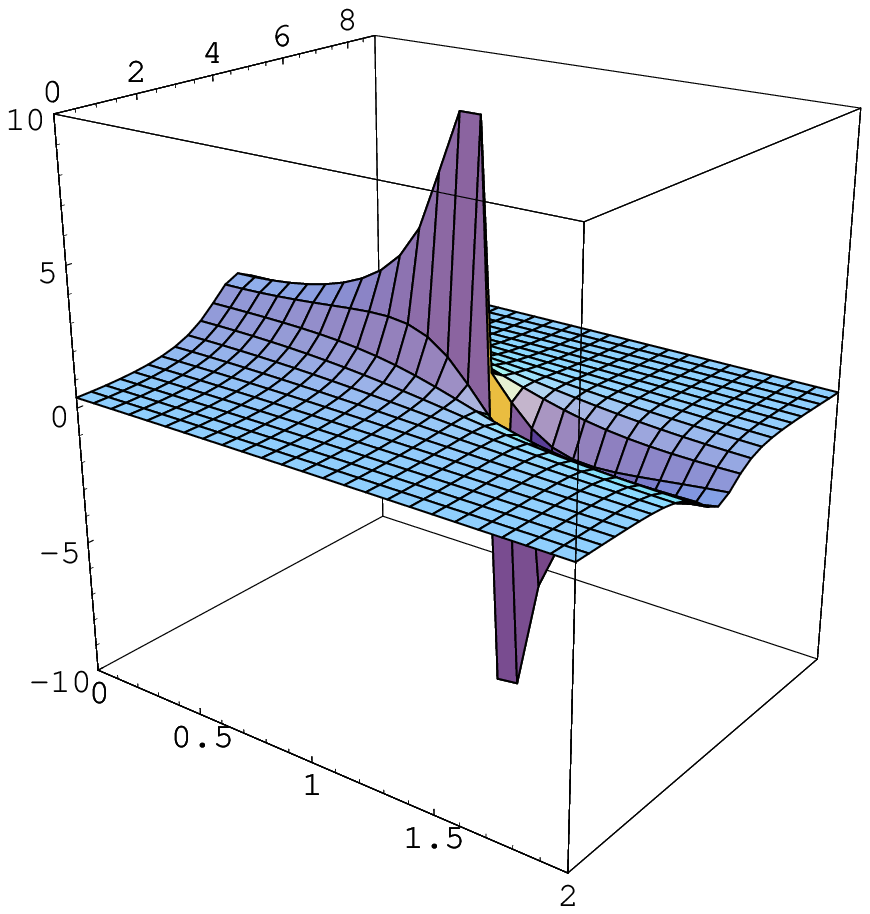}\hskip1.5cm
   \epsfysize=5.5cm  \epsffile{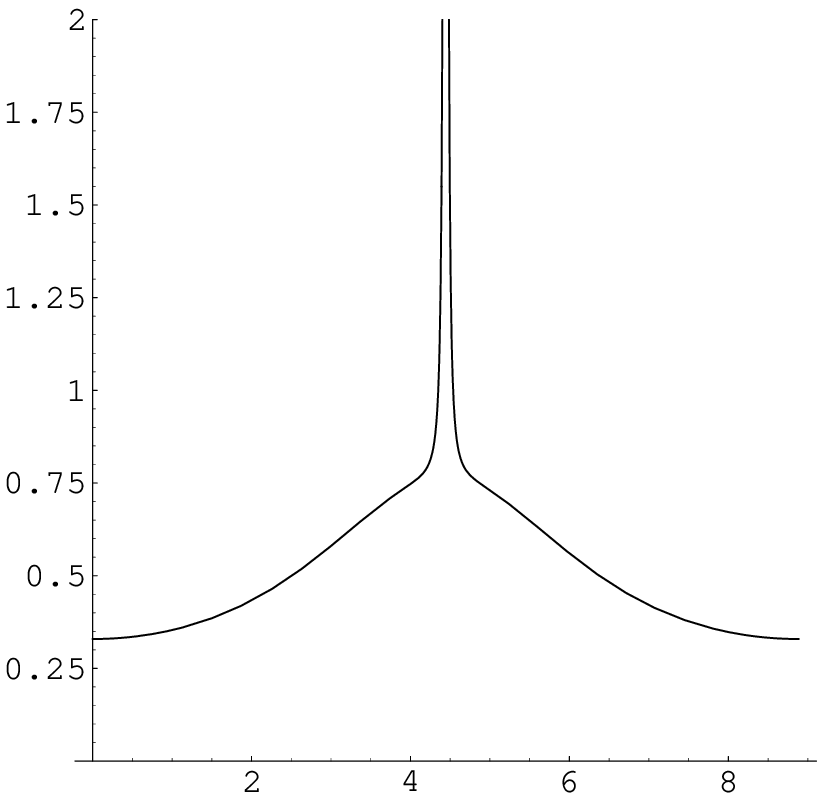}}
\caption{\sl Behavior of the energy density ($\propto {\rm
Re}f^+(x^0,x^1)$) (vertical axis in arbitrary unit) for $\lt=0.3$
in $x^0 = [0, 2]$ and $x^1 = [0, 2\sq\pi]$ in the left, and its
section just before the critical time $x_c^0\sim 1.116\cdots$
 in the right.} \label{enefig}
\end{figure}
 Utilizing the expressions Eq.(\ref{FG}), we obtain
\cite{seninhomo}
\be T_{00} = - T_{11}  &=& {\cal T}_{25} \cos^2
(\widetilde{\lambda} \pi/2) \, {\rm Re} f^+ (x^0, x^1), \nn T_{01}
= +T_{10}  &=& {\cal T}_{25} \cos^2(\widetilde{\lambda} \pi /2) \,
{\rm Im} f^+(x^0, x^1), \nn T_{ii} &=&-\cT_{25}||f^+(x^0,x^1)||^2
\qquad ({\rm for}~~~i=2,3,\cdots,25). \label{em-components} \ee
One readily sees that the energy-momentum tensor behaves as a sort
of impulse source to the closed string field equation. From
Eq.(\ref{ave}), the average energy is
\be \langle E \rangle (x^0) &:=& {1 \over 2\sq \pi}
\int_0^{2\sq\pi}dx^1 \, T_{00} (x^0,x^1) =\left\{
\begin{array}{ccc} 0 & {\rm for} & \vert x^0 \vert > x_c^0 \\
{\cal T}_{25}\cos^2(\lt\pi/2) & {\rm for} & \vert x^0 \vert < x_c^0
\end{array} \right. .
\nonumber \ee
It indicates that the energy suddenly disappears after the
critical time! This comes about as follows. As can be seen from
Eq.(\ref{realpart}) and Eq.(\ref{ave}), {\sl half} of the total
energy is squeezed at $x^1=\sq(2n+1)\pi$ ($n\in\Z$) and form the
$\delta$-function singularities as the critical time $x_c^0$ is
approached, while the other half is smoothly spread out along the
$x^1$-direction (See Fig.\ref{enefig}). After the critical time
$x_c^0$, the sign of the $\delta$-function flips (see
Eq.(\ref{realpart})) and cancels out contribution of the other
half that are smoothly spread out. The momentum density is zero on
the average, but develops a singular profile which may be
approximated by $\sim\tan(x^1/{2\sq})$ 
(See Fig.\ref{momfig}).
%

To gain a better picture concerning the origin of the singularity,
let us examine the energy-momentum conservation carefully.
Introduce for convenience complex coordinates $z = (x^0 + i x^1)$,
$\overline{z} = (x^0 - i x^1)$ so that
$\partial_0 = (\partial_z + \partial_{\overline z})$,
$\partial_1 = i (\partial_z - \partial_{\overline z})$.
The coefficient functions $f^\pm$ are analytic functions of $z,
\overline{z}$, respectively. From Eq.(\ref{em-components}), we
then find that
\be
\partial^a T_{a0} &=& - {\cal T}_{25} \cos^2(\wt{\lambda} \pi/2) \,
\left[ \partial_z f^-(\overline{z}) + \partial_{\overline{z}}
f^+(z) \right] \nn
\partial^a T_{a1} & = & \, i \, {\cal T}_{25} \cos^2( \wt{\lambda}
\pi/2) \, \left[ \partial_z f^-(\overline{z}) -
\partial_{\overline z} f^+(z) \right]. \label{ems} \ee

As alluded in the previous section, $f^+(z), f^-(\overline{z})$
have simple poles at $z = - \sqrt{2} \log(-\sin(\wt{\lambda}
\pi/2))$ and $\overline{z} = + \sqrt{2} \log(-\sin(\wt{\lambda}
\pi/2))$, respectively. Thus, the first relation in Eq.(\ref{ems})
is proportional to
$\sum_{n\in\Z}\delta(x^0 - x^0_c) \delta (x^1 - \sqrt{2}(2n+1)\pi)$,
invalidating the energy-momentum conservation. On the other
hand, the second relation in Eq.(\ref{ems}) vanishes, yielding the
correct force law.

Physical interpretation of such singular behavior is not clear to
us, so we shall primarily concentrate our consideration on the
early evolution before hitting the critical time, $i.e.$
$|x^0|<x_c^0$. As the total energy ought to be conserved, a
plausible possibility suggested in \cite{seninhomo} is that the
missing energy has escaped through the singularity to form an
array of codimension-one D-branes. We leave viability of this
scenario aside for future study. (See section 5 for further
discussion.)

\begin{figure}[tb]
   \vspace{0cm}
   \epsfysize=6cm
   \centerline{\epsffile{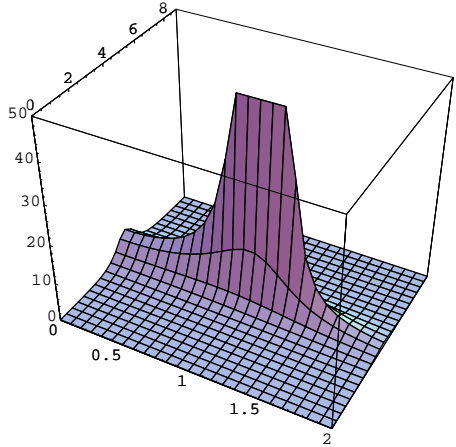}}
\caption{\sl Behavior of closed string tachyon and dilaton coupling
density ($\propto ||f^+(x^0,x^1)||$)
 (vertical axis in arbitrary unit) for $\lt=0.3$ in
$x^0 = [0,2]$ and $x^1 = [0,2\sq\pi]$.
} \label{absf}
\end{figure}

$\bullet$ \underline{tachyon coupling} \hfill\break
The tachyon coupling density is given by
\be \rho_{\rm tachyon} (x^0, x^1) = {\cal T}_{25} \vert \vert f^+
(x^0, x^1) \vert \vert^2, \nonumber \ee
and its behavior is plotted in Fig.\ref{absf}. The coupling
again develops a singularity at the critical time $x^0 = x^0_c$.
 {}From Eqs.(\ref{FG},\ref{coefficients}),
we also find that couplings to all higher
closed string modes with polarization along $i=2, 3, \cdots, 25$
directions behave the same way as the tachyon coupling.


%
%

$\bullet$ \underline{dilaton coupling} \hfill\break
As the part of the boundary state that couples to the sigma model
dilaton is
\be \left(c_0 + \overline{c}_0 \right) \left( c_{-1} c_1 +
\overline{c}_{-1} \overline{c}_1 \right) \vert 0 \rangle_{\rm gh},
\nonumber \ee
the dilaton coupling of the decaying D25-brane is given by
\be \rho_{\rm dilaton} (x^0, x^1) = {\cal T}_{25} \vert \vert
f^+(x^0, x^1) \vert \vert^2. \nonumber \ee
Its behavior is the same as that of the tachyon coupling and the
pressure $T_{ii} \,\, (i=2,3, \cdots, 25)$.

\begin{figure}[tb]
   \vspace{0cm}
   \centerline{
   \epsfysize=6cm  \epsffile{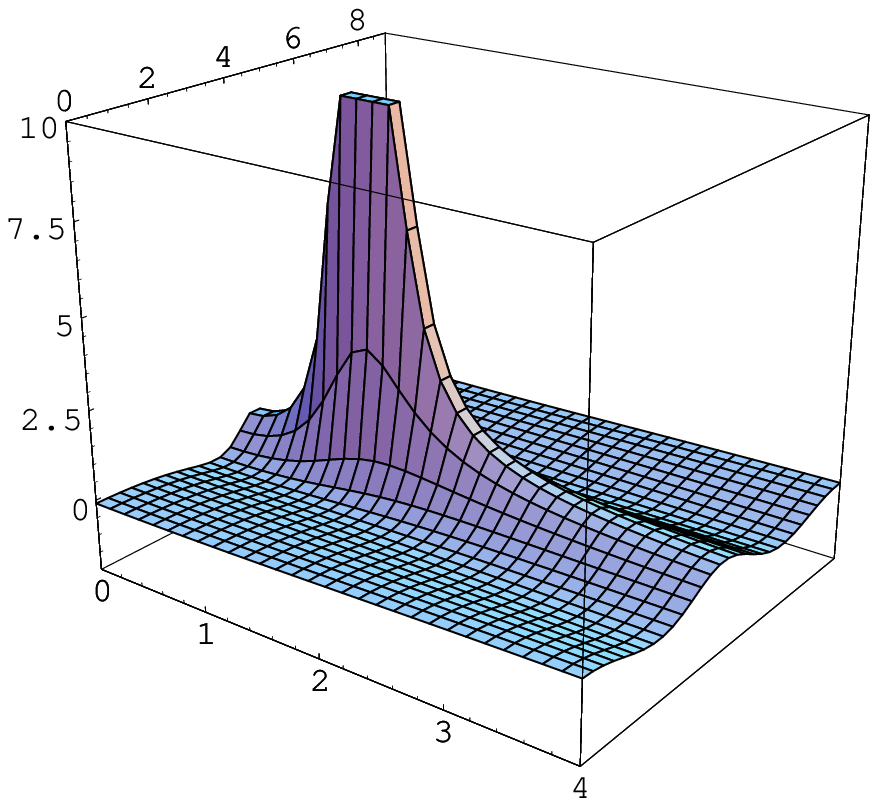} \hskip1cm
   \epsfysize=6cm  \epsffile{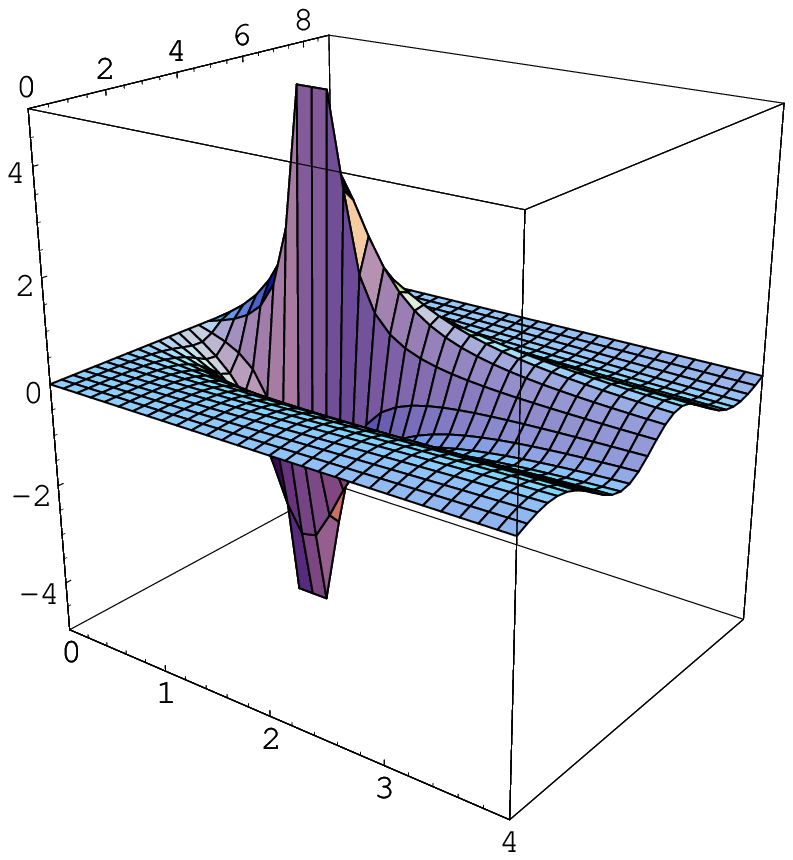}}
   \caption{\sl Behavior of $-{\rm Re}(f^+h_1^-)$ in the left
   and ${\rm Im}(f^+h_1^-)$ in the right
  (vertical axis in arbitrary unit) with $\lt=0.3$ in $x^0=[0,4]$
  and $x^1=[0,2\sq\pi]$.}
\label{hffig}
\end{figure}

$\bullet$ \underline{massive string mode coupling} \hfill\break
Coupling to massive closed string modes, as compared to tachyon
and massless string modes, entails certain new features. As is
seen from Eq.(\ref{coefficients}), these couplings are
distinguished from those for tachyon and massless modes in that,
in addition to $f^\pm(x^0, x^1)$, new coefficient functions
$h^\pm_1(x^0, x^1), h^\pm_2(x^0, x^1), h^\pm_3(x^0, x^1), \cdots$
are involved.
 We thus expect that time evolution of the massive
mode couplings would display qualitatively significant departure
from those of the tachyon and the massless state couplings.
Generally, they do not vanish at late time, though the energy
density vanishes soon after the system hits the singularity at the
critical time. As an example, let us consider $H_{ab}(x^0, x^1)$
couplings. They are governed by the real and the imaginary parts
of the coefficient functions $f^\pm$ and $h_1^\pm$:
\be H_{00} + i H_{01} = -H_{11}+i H_{10}=
 f^+ h_1^- \ . \nonumber
\ee
We plot the real and the imaginary parts of the function
\be f^+ h_1^- = 2 \cos^2 (\widetilde{\lambda} \pi/2) \left[1- 2
\sin (\widetilde{\lambda} \pi /2) \cosh {(x^0 - i x^1) \over
\sqrt{2}} \right] f^+ - \vert \vert f^+\vert\vert^2 \nonumber \ee
in Fig.\ref{hffig}.
The plots show that a singularity develops precisely at the
critical time $x^0=x^0_c$, where the couplings to tachyon and
massless modes diverged. This is again attributable to the fact
that all these couplings depend on the function $f^\pm$. On the
other hand, for the asymptotic behavior, the $H_{ab}$ coupling
differs from the couplings to tachyon and massless modes. At $x^0
\gg x^0_c$, we find that
\be f^+h_1^-\sim -2\cos^2(\lt\pi/2) e^{-\sq i x^1}, \nonumber \ee
exhibiting persistent modulation along the $x^1$-direction. Note
that, as compared to the early epoch $x^0 \ll x^0_c$, the
modulation wave-number has increased {\sl twice} asymptotically.

The coupling functions at higher massive levels would be governed
by two qualitatively different contributions. These coupling
functions are extractable from expanding the product of two $c=1$
boundary states Eq.(\ref{factorized}) in powers of closed string
oscillator levels. For the first massive level modes, we have seen
in Eq.(\ref{coefficients}) that the couplings are quadratic
products among the coefficient functions ($f^\pm, g^\pm, h^\pm_1,
h^\pm_2, h^\pm_3$) up to level-2. Likewise, for a generic massive
level mode, the coupling would be quadratic products among the
coefficient functions, now involving higher-level ones. Expanding
the products, we see that the coupling function comprises of two
kind of contributions: the one depending on $f^\pm$ (e.g. most of
Eq.(\ref{coefficients}) in the case of the first massive level
modes) and the one independent of $f^\pm$ (e.g. products among the
first parts of $g^\pm, h^\pm_1, h^\pm_2, h^\pm_3$ in the case of
the first massive level modes). As the first type of contributions
depends on $f^\pm$, there will always be a singularity at $x^0 =
x^0_c$. The second type of contributions is non-singular,
generally increasing as $x^0 \rightarrow \infty$, and hence is
qualitatively analogous to the behavior of the couplings for
homogeneous tachyon rolling.

A remark is in order. So far, we have focused primarily on
constructing the boundary state Eq.(\ref{expandedbs}) itself. It
might well be that some components of the boundary state
Eq.(\ref{expandedbs}) are BRST exact, viz. couple only to
off-shell or unphysical closed string states. In that case,
recalling that the boundary state Eq.(\ref{expandedbs}) acts as a
source to the closed string field equation Eq.(\ref{sfteq}), the
BRST exact components would not be contributing to on-shell closed
string processes, such as closed string radiation out of decaying
D-brane. Whether there indeed exists BRST exact components and, if
so, which modes (and their time-dependent coefficient functions)
belong to them are pertinent questions to be examined in case one
needs to study on-shell processes. In this work, we will leave the
issue aside for future study.

\section{Turing on Electric and Magnetic Fields}
With the physics motivation addressed in section 1, we extend the
analysis of section 2 and construct a boundary state for an
unstable D-brane with electric and magnetic fields turned on.

\subsection{Rolling of Modulated Tachyon under Electric Flux}
We shall now consider turning on a {\sl constant} electric field
on the D25-brane world-volume, and study rolling of the spatially
modulated tachyon field. As is well-known, the electric field
induces fundamental string constituents on the D25-brane
world-volume. If the tachyon field were spatially homogeneous, the
induced string charge density (which is given by the electric
displacement field) would be homogeneous over the D25-brane
world-volume as well. On the other hand, if the tachyon field were
spatially modulated, the induced string charge density would be
modulated accordingly and are driven to evolve as the tachyon
rolls down to the potential minimum. In such a circumstance, an
interesting and potentially important question is whether the
displacement field flux can possibly get squeezed into (an array
of) thin flux tubes. If so, adding spatial modulation on a
decaying D-brane would be a new way of manufacturing a
macroscopically large, fundamental closed string!

With such motivation, consider turning on the world-volume
electric field along $2$-direction $e=F_{02}$.
With tachyon modulation along
$1$-direction, this amounts to a fluid of fundamental strings,
whose macroscopic configuration is stretched along $2$-direction
and charge density varies across $1$-direction. As prescribed in
the work \cite{reysugimoto}, the corresponding boundary state is
obtainable by a sequence of T-dual map, boost by $e$, and inverse
T-dual map, all along the $2$-direction. Denote the new
coordinates as $y^a$'s and the new string oscillators as $\beta^a_n,
\overline{\beta}^a_n$'s. Starting from Eq.(\ref{expandedbs}), the
new boundary state would again be expressible as:
\be \vert B \rangle_{T, e} &=& F^e(y^0, y^1) \vert 0 \rangle \nn
&+& G^e_{ab}(y^0, y^1) \beta^a_{-1} \overline{\beta}^b_{-1} \vert
0 \rangle + {1 \over 2} H^e_{ab} (y^0, y^1) \beta^a_{-2}
\overline{\beta}^b_{-2} \vert 0 \rangle \nn
&+& {1\over 4} I^e_{abcd}(y^0,y^1) \beta^a_{-1} \beta^b_{-1}
\overline{\beta}^c_{-1} \overline{\beta}^d_{-1} \vert 0 \rangle
\nn &+& {i \over 2} J_{abc}^e(y^0, y^1) \left( \beta^a_{-1}
\beta^b_{-1} \overline{\beta}^c_{-2} + \beta^c_{-2}
\overline{\beta}_{-1}^a \overline{\beta}_{-1}^b \right) \vert 0
\rangle + \cdots, \nonumber \ee
where
\be F^e(y^0, y^1) &=& \gamma^{-1} F(\gamma^{-1} y^0, y^1) \nn
G^e_{ab}(y^0, y^1) &=& \gamma^{-1} \left( {}^t \Lambda^{-1} G
\Lambda \right)_{ab} (\gamma^{-1} y^0, y^1) \nn
H^e_{ab}(y^0, y^1) &=& \gamma^{-1} \left( {}^t \Lambda^{-1} H
\Lambda \right)_{ab} (\gamma^{-1} y^0, y^1) \nn
I^e_{abcd}(y^0, y^1) &=& \gamma^{-1} \left( {}^t \Lambda^{-1}
\left( {}^t \Lambda^{-1} I \Lambda \right)_{ac} \Lambda
\right)_{bd} (\gamma^{-1} y^0, y^1) \nonumber \ee
etc., and
\be \Lambda = \left( \begin{array}{ccccc}
\gamma & 0 & \gamma e & &  \\
0 & 1 & 0 & &  \\
\gamma e & 0 & \gamma & & \\
&&&1 &   \\
&&&& \ddots \end{array} \right),~~~\gamma=\frac{1}{\sqrt{1-e^2}}.
\nonumber \ee
Explicitly, the new coefficient functions are given as
\be F^e(y^0, y^1) &=& \gamma^{-1} F (\gamma^{-1} y^0, y^1) \nn &=&
\gamma^{-1} \vert\vert f^+ (\gamma^{-1} y^0, y^1)\vert\vert^2, \nn
G^e_{00}(y^0, y^1) &=& \gamma \left[G_{00}(\gamma^{-1} y^0,y^1) -
e^2 G_{22}(\gamma^{-1} y^0, y^1) \right] \nn &=& - \gamma^{-1}
\vert\vert f^+ (\gamma^{-1} y^0, y^1)\vert\vert^2 + \, 2 \gamma
 \cos^2(\lt \pi /2)\, {\rm Re} f^+(\gamma^{-1}y^0, y^1),
\nn
G^e_{11} (y^0, y^1) &=& \gamma^{-1} G_{11}(\gamma^{-1} y^0,
y^1)\nn &=&+\gamma^{-1} \vert \vert f^+(\gamma^{-1} y^0, y^1)
\vert\vert^2 - 2 \gamma^{-1}\cos^2(\lt \pi/2) \, {\rm Re}
f^+(\gamma^{-1} y^0, y^1), \nn
 G^e_{22}(y^0, y^1) &=& \gamma
\left[G_{22}(\gamma^{-1} y^0, y^1) - e^2 G_{00}(\gamma^{-1} y^0,
y^1) \right] \nn &=& - \gamma^{-1} \vert\vert f^+(\gamma^{-1}y^0,
y^1)\vert\vert^2 - 2 \gamma e^2 \cos^2(\lt\pi/2) \, {\rm Re}
f^+(\gamma^{-1}y^0, y^1), \nn
G^e_{ii}(y^0, y^1) &=& \gamma^{-1} G_{ii} (\gamma^{-1} y^0, y^1)\nn
&=& -\gamma^{-1} \vert \vert f^+(\gamma^{-1} y^0, y^1) \vert\vert^2
\qquad ({\rm for}~~~ i = 3, \cdots, 25),
 \nn
G^e_{02}(y^0, y^1) = - G^e_{20}(y^0, y^1) &=& \gamma e \left[
G_{00} (\gamma^{-1} y^0, y^1) - G_{22}(\gamma^{-1} y^0, y^1)
\right] \nn &=& 2 \gamma e \cos^2(\lt \pi/2 ) \, {\rm Re}
f^+(\gamma^{-1} y^0, y^1), \nn
 G^e_{01}(y^0, y^1) = + G^e_{10}(y^0,
y^1) &=& \, G_{01}(\gamma^{-1} y^0, y^1) = \,\, 2
\cos^2(\lt\pi/2) \, {\rm Im} f^+(\gamma^{-1}y^0, y^1), \nonumber
\nn
G^e_{12}(y^0, y^1) = - G^e_{21}(y^0, y^1) &=& e G_{01}(\gamma^{-1}
y^0, y^1) = 2e\cos^2(\lt\pi/2) \, {\rm Im} f^+(\gamma^{-1} y^0,
y^1) \nonumber \ee
for the tachyon and the massless level modes, and
\be H_{00}^e(y^0, y^1) &=& \gamma \left[ H_{00} - e^2 H_{22}
\right] = +\gamma \left[{\rm Re} (f^+ h_1^-) + e^2 \vert \vert f^+
\vert \vert^2 \right](\gamma^{-1} y^0, y^1), \nn
H^e_{11}(y^0, y^1) &=& \gamma^{-1} H_{11} = - \gamma^{-1} {\rm Re}
(f^+ h_1^-)(\gamma^{-1} y^0, y^1), \nn
H^e_{22} (y^0, y^1) &=& \gamma \left[ H_{22} - e^2 H_{00} \right]
= -\gamma \left[e^2 {\rm Re} (f^+ h_1^-) + \vert \vert f^+ \vert
\vert^2 \right] (\gamma^{-1} y^0, y^1),\nn
H_{ii}^e (y^0, y^1) &=& \gamma^{-1} H_{ii} =
-\gamma^{-1} \vert \vert f^+ (\gamma^{-1} y^0, y^1)\vert \vert^2
\qquad ({\rm for}~~~i = 3, \cdots, 25), \nn
H^e_{01}(y^0, y^1) = + H^e_{10}(y^0, y^1) &=& \, H_{01} = \,\,
{\rm Im} (f^+ h_1^-)(\gamma^{-1} y^0, y^1), \nn
H^e_{12}(y^0, y^1) = - H^e_{21}(y^0, y^1) &=& e H_{01} = e\, {\rm
Im} (f^+ h_1^-)(\gamma^{-1} y^0, y^1), \nn
H^e_{02}(y^0, y^1) = - H^e_{20}(y^0, y^1) &=& e \gamma
\left[H_{00} - H_{22} \right] = e \gamma \left[ {\rm Re} (f^+
h_1^-) + \vert \vert f^+ \vert \vert^2 \right](\gamma^{-1} y^0,
y^1) \nonumber \ee
for the coupling to the first massive level modes.

\subsection{Coupling to Closed String Modes}
Coupling to the closed string modes are readily obtained from the
boundary state, as we did in section 2.

$\bullet$ \underline{tachyon and dilaton couplings}
\hfill\break
For tachyon and dilaton fields, the couplings are
given by
\be \rho^e_{\rm tachyon}(y^0, y^1) = \rho^e_{\rm dilaton}(y^0,y^1)
= {\cal T}_{25} \gamma^{-1}
 \vert \vert f^+ (\gamma^{-1} y^0,y^1)\vert \vert^2 \ . \nonumber \ee
We see that, as the electric field is turned on along
$x^2$-direction, the rolling is time-dilated and the spatial
modulation along $x^1$-direction remains unaffected. In the
critical limit, $e \rightarrow 1$, these couplings vanish because
the overall Born-Infeld factor $\gamma^{-1}$ dilutes the coupling
density.

$\bullet$ \underline{graviton coupling} \hfill\break
 The
energy-momentum tensor is given in terms of the coupling functions
by \be T^e_{ab}(y^0, y^1) = {{\cal T}_{25} \over 2} \left(
G^e_{(ab)}(y^0, y^1) - \eta_{ab} F^e (y^0, y^1) \right), \nonumber
\ee
so we obtain
\be T^e_{00}(y^0, y^1) &=& + \, {\cal T}_{25} \gamma \,
\cos^2(\lt\pi/2 ) \, {\rm Re} f^+ (\gamma^{-1} y^0, y^1),
\nn
T^e_{01} (y^0, y^1) &=& +{\cal T}_{25} \cos^2 (\widetilde{\lambda}
\pi/2)\, {\rm Im} f^+ (\gamma^{-1} y^0, y^1), \nn
T^e_{11} (y^0, y^1) &=& - {\cal T}_{25} \gamma^{-1}
\cos^2(\widetilde{\lambda} \pi/2)\, {\rm Re} f^+ (\gamma^{-1} y^0,
y^1), \nn
T^e_{22} (y^0, y^1) &=& - {\cal T}_{25} e^2 \gamma \cos^2
(\widetilde{\lambda} \pi/2) \, {\rm Re} f^+ (\gamma^{-1} y^0, y^1)
- {\cal T}_{25} \gamma^{-1}
|| f^+(\gamma^{-1} y^0, y^1) ||^2, \nn
T^e_{ii} (y^0, y^1) &=& -{\cal T}_{25}\gamma^{-1}
||f^+(\gamma^{-1} y^0, y^1)||^2 \qquad
({\rm for}~~~ i = 3, \cdots, 25).
\label{Te}
\ee
Let us compare Eq.(\ref{Te}) with the energy-momentum tensor
Eq.(\ref{em-components}) at zero electric flux. Apart from the
expected Born-Infeld time dilation and Lorentz contraction, we see
that the pressure along the electric field direction receives a
new contribution, proportional to the energy density. This part is
attributed to the fundamental string constituents.

$\bullet$ \underline{Kalb-Ramond coupling} \hfill\break
At nonzero electric flux, coupling to the Kalb-Ramond field
\be {\cal Q}^e_{ab}(y^0, y^1) &:=& {{\cal T}_{25} \over 2}
G^e_{[ab]}(y^0, y^1), \nonumber \ee
is newly induced, whose non-zero components are
\be
 {\cal Q}_{[02]}(y^0, y^1) &=& {\cal T}_{25} \gamma e \cos^2(
\widetilde{\lambda} \pi/2) \, {\rm Re} f^+(\gamma^{-1} y^0, y^1),
\nn
{\cal Q}_{[12]} (y^0, y^1) &=& {\cal T}_{25} \, e  \cos^2
(\widetilde{\lambda} \pi/2) \, {\rm Im} f^+(\gamma^{-1} y^0, y^1).
\label{Qe}
\ee
As anticipated, the induced Kalb-Ramond coupling is proportional
to the nonzero electric field, thus making up the decaying
D25-brane to carry a fundamental string charge. Note that the
charge density of the fundamental string stretched along
2-direction is given by ${\cal Q}_{[02]}$, and it is
proportional to the energy density in Eq.(\ref{Te}) as
\begin{eqnarray}
{\cal Q}_{[02]}(y^0, y^1)=e \, T_{00}^e(y^0, y^1) \ . \label{Q=eT}
\end{eqnarray}
Likewise, the current density of the fundamental string ${\cal
Q}_{[12]}$ is proportional to the momentum density in
Eq.(\ref{Te}) as
\be {\cal Q}_{[12]}(y^0, y^1) = e \, T_{01}^e(y^0, y^1) \ .
\label{J=eP} \ee
Therefore, just as the behavior of energy density explained in
section \ref{coupling}, {\sl half} of the fundamental string
charge density would squeeze to form (an array of) localized
$\delta$-function profile as the time $y^0$ approaches the
critical time $y^0_c$. The off-diagonal (01)-component of the
energy-momentum tensor and (12)-component of the Kalb-Ramond
coupling imply that the fundamental string density is actually
flowing along the $1$-direction! This is to be contrasted with the
situation of homogeneous tachyon rolling, where turning on
electric field $e$ induces a fundamental string fluid on the
decaying D-brane world-volume, but the fluid is {\sl at rest}. As
shown in \cite{reysugimoto,kim3kwon}, turning on magnetic field
$b$ perpendicular to the electric field lets the fluid to flow
rigidly. We now find that modulated tachyon field can also trigger
the fundamental string gas to flow. Moreover, the flow velocity
varies spatially, albeit it is determined by the modulation of the
rolling tachyon itself.

$\bullet$ \underline{BPS Limit} \hfill\break
 Now that the electric field $e$ is restricted to $|e| \le 1$,
Eqs.(\ref{Q=eT}, \ref{J=eP}) imply the following BPS-like
inequalities
\be T^e_{00} \ge \vert {\cal Q}^e_{[02]} \vert \qquad {\rm and}
\qquad |T^e_{01}| \ge |{\cal Q}^e_{[12]}| \ , \nonumber \ee
which are saturated precisely at the limit $|e| \rightarrow 1$.

One important effect of turning on the electric field is the time
dilation by the $\gamma^{-1}$ factor: $x^0 \rightarrow \gamma^{-1}
y^0$. Therefore, the critical time is now dilated to
\be y^0_c = \gamma\sq\log\left({1 \over
\vert\sin(\lt\pi/2)\vert}\right) = \gamma x^0_c
. \nonumber \ee
If we take the extremal limit
\be e \rightarrow 1 \qquad {\rm and} \qquad \widetilde{\lambda}
\rightarrow 1 \nonumber \ee
%
while holding the average energy $\langle E \rangle =\gamma {\cal
T}_{25}\cos^2(\lt\pi/2)$ fixed, we find that the critical time is
situated at a finite value:
\be y^0_c \rightarrow {E \over \sq\,{\cal T}_{25}} \ ,
\nonumber \ee
indicating that the relaxation time scale grows with the total
energy one starts with.

In the extremal limit, the real part of $f^\pm$ becomes
\be
{\rm Re} f^\pm(\gamma^{-1}y^0,y^1)
\rightarrow \left\{
\begin{array}{ccc}
 2\sq\pi \sum_{n\in\Z}\delta\left(y^1 -\sq (2n+1) \pi\right) &
{\rm for} & \vert y^0 \vert < y^0_c,\\
0 & {\rm for} & \vert y^0 \vert > y^0_c. \end{array} \right.
\nonumber \ee
Thus, prior to reaching the critical time $y^0_c$, non-vanishing
components of the energy-momentum tensor and the Kalb-Ramond
tensor current density become
\begin{eqnarray}
T^e_{00}=-T^e_{22}=|{\cal Q}_{02}|=2\sq\pi E\sum_{n\in\Z}\delta\l(
y^1-\sq(2n+1)\pi\r) \ . \label{BPS}
\end{eqnarray}
This is the same relation as that saturated
by an array of BPS fundamental strings stretched along the
$y^2$-direction, except that we are now considering homogeneous
distribution of the string fluid along the transverse
$y^3,\cdots,y^{25}$-directions.
It thus indicates that the fundamental string constituents are
confined to co-dimension one hypersurfaces located at $y^1 =
\sqrt{2} (2n+1) \pi$.

Note, however, that this extremal limit is not a smooth limit for
the full boundary state. As pointed out in
\cite{MuSen,reysugimoto} for the homogeneous case, the
coefficients in the boundary state for higher-level massive modes
with more than three oscillators for 0- or 2-directions are
weighted with extra $\gamma$ factors, and hence will diverge in
the extremal limit. In light of the remark at the end of section
2, a possible way out is that (part of) the non-smooth modes
belong to the BRST exact class.

\subsection{The Effect of Magnetic Field}
Physics-wise, it is also of interest to turn on magnetic field on
the decaying D-brane world-volume. The world-volume gauge field is
combined with pull-back of the Kalb-Ramond field $X^* B_{ab}$ into
a gauge-invariant combination $(F_{ab} + X^* B_{ab})$. It thus
implies that nonzero magnetic field induces {\sl current} density
of the fundamental string on the decaying D-brane world-volume.
Here, we examine how the fundamental string constituents are
affected by the magnetic field. We also consider the case with
pure magnetic field. In this case, the total fundamental string
charge is zero, but, nevertheless we find that polarized
fundamental string charge density is induced, and the system
evolves to a configuration with an array of fundamental string -
anti-fundamental string pairs (i.e. an array of fundamental
strings of alternating orientations). We will see that, again,
spatial modulation of the rolling tachyon field plays a prominent
role.

Following the prescription given in the previous work
\cite{reysugimoto}, we find that the boundary state is
\be
\vert B \rangle_{T, e+b} &=& F^{e + b}(y^0, y^1) \vert 0
\rangle \nn
&+& G^{e+b}_{ab}(y^0, y^1) \, \beta^a_{-1} \overline{\beta}^b_{-1}
\vert 0 \rangle + {1 \over 2} H_{ab}^{e+b}(y^0, y^1) \,
\beta^a_{-2} \overline{\beta}^b_{-2} \vert 0 \rangle \nn &+& {1
\over 4} I_{abcd}^{e+b} (y^0, y^1) \, \beta^a_{-1} \beta^b_{-1}
\overline{\beta}^c_{-1} \overline{\beta}_{-1} \vert 0 \rangle + {i
\over 2} J_{abc}(y^0, y^1) \left( \beta^a_{-1} \beta^b_{-1}
\overline{\beta}^c_{-2} + \beta^c_{-2} \overline{\beta}^a_{-1}
\overline{\beta}^b_{-1} \right) \vert 0 \rangle \nn &+& \cdots \ ,
\nonumber \ee
where the coefficient functions are obtained as
\be F^{e+b}(y^0, y^1) &=& \gamma^{-1} \widetilde{\gamma}^{-1} F
(\gamma^{-1} y^0, \widetilde{\gamma}^{-1} y^1 - be
\widetilde{\gamma} y^0), \nn
G^{e+b}_{ab}(y^0, y^1) &=& \gamma^{-1} \widetilde{\gamma}^{-1}
\left( {}^t(\Omega \Lambda)^{-1} G  \, \Lambda \Omega \right)_{ab}
(\gamma^{-1}y^0, \widetilde{\gamma}^{-1} y^1 - be
\widetilde{\gamma} y^0), \nn
H^{e+b }_{ab}(y^0, y^1) &=& \gamma^{-1} \widetilde{\gamma}^{-1}
\left( {}^t(\Omega \Lambda)^{-1} H  \, \Lambda \Omega \right)_{ab}
(\gamma^{-1}y^0, \widetilde{\gamma}^{-1} y^1 - be
\widetilde{\gamma} y^0), \nn
I^{e+b}_{abcd}(y^0, y^1) &=& \gamma^{-1} \widetilde{\gamma}^{-1}
\left( {}^t(\Omega \Lambda)^{-1} \left( {}^t (\Omega \Lambda)^{-1}
I \, \Lambda \Omega \right)_{bc} \Lambda \Omega
\right)_{ad}(\gamma^{-1}y^0, \widetilde{\gamma}^{-1} y^1 - be
\widetilde{\gamma} y^0), \nonumber
\ee
etc., in terms of transformation matrices
\begin{eqnarray}
\Lambda=
\left(
\begin{array}{ccccc}
\gamma&0&\gamma e'\\
0&1&0\\
\gamma e'&0&\gamma\\
&&&1\\
&&&&\ddots
\end{array}
\right) \qquad {\rm and} \qquad  \Omega= \left(
\begin{array}{ccccc}
1&0&0\\
0&\wt\gamma&-\wt\gamma b\\
0&+\wt\gamma b&\wt\gamma\\
&&&1\\
&&&&\ddots
\end{array}
\right) \ , \nonumber
\end{eqnarray}
and
\begin{eqnarray}
\gamma=\frac{1}{\sqrt{1-{e'}^2}}~~{\rm with}~~
e'=\frac{e}{\sqrt{1+b^2}} \qquad {\rm and} \qquad
\wt\gamma=\frac{1}{\sqrt{1+b^2}}. \nonumber
\end{eqnarray}
We then find that the nonzero components of the energy-momentum
tensor are
\be T_{00}^{e+b}&=& + \cT_{25} \gamma \widetilde{\gamma} (1 +
b^2)\cos^2(\lt\pi/2)\, {\rm Re}f^+,\nn
T_{01}^{e+b} &=& + \cT_{25}\cos^2(\lt\pi/2)\left[{\rm Im}f^+ +
\gamma \wt\gamma eb\,{\rm Re}f^+ \right], \nn
T_{11}^{e+b} &=& - \cT_{25} \cos^2(\lt\pi/2)\left[ \gamma
\widetilde{\gamma} (1 - e^2)\,{\rm Re}f^+ -{2eb \over (1 +
b^2)}\,{\rm Im} f^+ \right],\nn
 T^{e+b}_{22} &=&  + \cT_{25} \cos^2(\lt\pi/2)
\left[ \gamma \widetilde{\gamma} (b^2 - e^2) \,{\rm Re}f^+ + {2 eb
\over (1 + b^2)}\,{\rm Im}f^+ \right] - \cT_{25}
\gamma^{-1}\wt{\gamma}^{-1}|| f^+||^2,\nn
 T_{ii}^{e+b} &=& -\cT_{25}
\gamma^{-1} \wt{\gamma}^{-1} ||f^+ ||^2 \qquad \qquad ({\rm
for}~~~i = 3, \cdots, 25), \nonumber \ee
while the nonzero components of the Kalb-Ramond coupling tensor
are
\be {\cal Q}_{[02]}^{e+b} &=& \cT_{25}\cos^2(\lt\pi/2) \left[
\gamma \wt{\gamma} e \,{\rm Re}f^+ - b \,{\rm Im}f^+ \right], \nn
{\cal Q}_{[12]}^{e+b} &=& \cT_{25} \cos^2(\lt\pi/2) \left[
 e \left({ 1 - b^2 \over 1+ b^2} \right)\,{\rm Im}f^+
+\gamma\wt{\gamma} b \,{\rm Re}f^+  \right], \nonumber \ee
where $f^\pm := f^\pm(\gamma^{-1} y^0,\wt{\gamma}^{-1} y^1+
eb\wt{\gamma} y^0) $. One can check that the conservation
equations $\del^a T_{ab}^{e+b}=0$ and $\del^a \cQ_{ab}^{e+b}=0$ are
satisfied. Couplings to other closed string modes are also
obtainable analogously from the prescription given in
\cite{reysugimoto}. Before proceeding further, we will contrast
the above results against those obtained in the previous sections.

First, the spatial coordinate $x^1$ is now replaced by
$(\wt\gamma^{-1}y^1+eb\wt\gamma y^0)$ in the functions $f^\pm$, so
the whole system would be moving in the $y^1$-direction with a
constant velocity
\begin{eqnarray}
V:=\frac{eb}{1+b^2} \quad \le 1 \ . \label{speed}
\end{eqnarray}
Indeed, prior to the critical time $y^0_c$, the average momentum
\begin{eqnarray}
\langle P \rangle
:=\frac{1}{2\sq\pi}\int_0^{2\sq\pi}dy^1\,T_{01}^{e+b}(y^0,y^1)=
\gamma\wt\gamma eb \, \cT_{25}\cos^2(\lt\pi/2) \nonumber
\end{eqnarray}
and the average energy
\begin{eqnarray}
\langle E\rangle
:=\frac{1}{2\sq\pi}\int_0^{2\sq\pi}dy^1\,T_{00}^{e+b}(y^0,y^1)=\gamma\wt\gamma
(1+b^2) \, \cT_{25}\cos^2(\lt\pi/2) \nonumber
\end{eqnarray}
are nonzero, and the velocity of the energy flow is computed
as $\VEV{P}/\VEV{E}=eb/(1+b^2)$. This is consistent
with Eq.(\ref{speed}).

Second, we note that the (01)-, (11)-, (22)-components of the
energy-momentum tensor are now composed of both real and imaginary
parts of $f^\pm$. Recall that real and imaginary parts of $f^\pm$
are even and odd functions, respectively, of the {\sl comoving}
spatial coordinate $(y^1+Vy^0)$, and develop a singularity at the
critical time $y^0_c$. This gives rise to several effects. For
(01)-component of the energy-momentum tensor, Poynting field
momentum of the electromagnetic fields (proportional to $eb$ in
magnitude) adds up into an inhomogeneous distribution along
$y^1$-direction, and accumulates to an array of delta functions as
the critical time $y^0_c$ is approached. For the (11)- and
(22)-components of the energy-momentum tensor, the electromagnetic
fields induce additional contribution proportional to ${\rm Im}
f^+$ whose magnitude is determined by the velocity $V$.

Third, [02]- and [12]-components of the Kalb-Ramond coupling
tensor indicate that the fundamental string charge and current
densities are induced not only by the electric field $e$ but also
separately by the magnetic field $b$. We are thus led to a
surprising observation that, even without the electric field, a
decaying D25-brane can carry fundamental string constituents
induced by the modulated tachyon and the magnetic flux! In this
case, as real and imaginary parts of $f^\pm$ are even and odd
functions of $(y^1+V y^0)$, the net charge would be zero while the
net current is nonzero --- the system is a sort of plasma of the
fundamental strings (consisting of strings of alternating
orientations).

Of particular interest is the extremal limit. Take
\be e \rightarrow \sqrt{1+b^2}
\qquad {\rm and} \qquad \widetilde{\lambda} \rightarrow 1,
\nonumber \ee
while keeping the average energy $\VEV{E}$ held finite. Then, the terms
which survive in this limit are
\be T_{00}^{e+b}(y^0, y^1) &=& \, \, \Gamma^2 \, \left( 2\sq\pi
\wh E \, \delta (y^0, y^1) \right) \, = - \, \Gamma^2 \,
T_{22}^{e+b}(y^0, y^1) \nn
T_{01}^{e+b}(y^0, y^1) &=&\Gamma^2 \, V \left( 2\sq\pi \wh E \,
\delta(y^0, y^1) \right) \, = \, V \, T_{00}^{e+b}(y^0, y^1) \nn
T_{11}^{e+b}(y^0, y^1) &=& \Gamma^2 V^2 \left( 2\sq\pi \wh E \,
\delta (y^0, y^1) \right) = V^2 T_{00}^{e+b}(y^0, y^1) \nn
T^{e+b}_{22}(y^0, y^1) &=&-2\sq\pi \wh E\, \delta (y^0, y^1) \nn
{\cal Q}_{[02]}^{e+b}(y^0, y^1) &=& 2\sq\pi\wh E\,\Gamma\,\delta
(y^0, y^1)\nn {\cal Q}_{[12]}^{e+b}(y^0, y^1) &=& 2\sq\pi\wh
EV\Gamma\,\delta (y^0, y^1) = V {\cal Q}^{e+b}_{[02]}(y^0, y^1) \
. \nonumber \ee
Here, $V=b/\sqrt{1+b^2}$ is the velocity Eq.(\ref{speed}) in the
extremal limit, $\Gamma=1/\sqrt{1-V^2}=\sqrt{1+b^2}$ is the
corresponding gamma factor, $\wh E=\VEV{E}/\Gamma^2$ is the average
energy in the rest frame, and
\begin{eqnarray}
\delta(y^0, y^1) :=\sum_{n\in\Z}\delta\l( \Gamma(y^1+V y^0)
-\sq(2n+1)\pi \r) \nonumber
\end{eqnarray}
is an array of the delta functions. As such, each component of the
energy-momentum and the fundamental string current tensors have a
constant magnitude. Evidently, if we observe the system in the
rest frame, the results are in agreement with the BPS equality,
Eq.(\ref{BPS}). Therefore, the system is viewed most transparently
as an array of BPS fundamental strings stretched along the
$y^2$-direction, but now moving in $y^1$-direction uniformly with
a constant velocity $V$.

It would also be of interest to consider the case with large
magnetic field to learn the effect of pure magnetic field. If we
set $e=0$ and take $b\ra\infty$, the leading terms for the
energy-momentum tensor and the Kalb-Ramond coupling tensor are
\begin{eqnarray}
T^{b}_{00}(y^0, y^1) &\simeq& +\cT_{25} \cos^2(\lt\pi/2)\,b\,{\rm
Re}f^+(y^0, y^1)\nn
T^{b}_{01}(y^0, y^1)&\simeq& +\cT_{25} \cos^2(\lt\pi/2) \,{\rm
Im}f^+(y^0, y^1)\nn
T^{b}_{11}(y^0, y^1) &\simeq& {\cal O}(b^{-1})\nn
T^{b}_{22}(y^0, y^1)&\simeq& +\cT_{25}\cos^2(\lt\pi/2)\,b\,{\rm
Re}f^+(y^0, y^1) -\cT_{25}\, b\,||f^+||(y^0, y^1)\nn
T^{b}_{ii}(y^0, y^1)&\simeq& -\cT_{25}\, b\,||f^+||(y^0, y^1)\nn
\cQ^{b}_{[02]}(y^0, y^1)&\simeq&-\cT_{25}\cos^2(\lt\pi/2)\,b\,{\rm
Im}f^+(y^0, y^1)\nn
\cQ^{b}_{[12]}(y^0, y^1)&\simeq&+\cT_{25} \cos^2(\lt\pi/2)\, {\rm
Re}f^+(y^0, y^1),
\end{eqnarray}
with $f^\pm(y^0, y^1) = f^\pm(y^0,by^1)$. Now, the period in the
$y^1$-direction is shortened to $2\sq\pi/b$. The energy and
momentum densities are proportional to ${\rm Re}f^+$ and ${\rm
Im}f^+$, respectively, which is qualitatively similar to the
previous results Eq.(\ref{em-components}) or Eq.(\ref{Te}). On the
other hand, the fundamental string charge and current densities
are now proportional to ${\rm Im}f^+$ and ${\rm Re}f^+$,
respectively, which is opposite to what we have found for the case
with pure electric field Eq.(\ref{Qe}). The behavior of the
fundamental string charge and current densities can be read from
Fig.\ref{momfig} and Fig.\ref{enefig}, respectively. Since ${\rm
Im}f^+=0$ at $y^0=0$, the charge density is initially zero
everywhere, but then starts to get polarized as the tachyon rolls
down the potential hill.

Intuitively, such a behavior can be understood as follows. As one
can readily see from the gauge invariance $X^*B \rightarrow X^*B +
\d \Lambda_1$, $F \rightarrow F - \d \Lambda_1$, the magnetic
field on a D-brane world-volume induces fundamental string {\sl
current} density.
 However, when the tachyon field is at the
bottom of its potential and the D-brane disappears, there is
nowhere for the string current to flow. Therefore, the region of
the D-brane world-volume where the tachyon is near the bottom of
the potential acts just like an insulator for an electric current.
In our situation, as the modulated tachyon evolves, the regions
around $y^1=2\sq n\pi/b$ ($n\in\Z$) behave as insulators while
those near $y^1=\sq(2n+1)\pi/b$ ($n\in\Z$) behave as conductors
for the string current. Thus, our system is analogous to a series
of capacitors injected with a non-zero electric current, in which
electric charge starts to accumulate at the edges of the
conductors. The polarization of the fundamental string {\sl
charge} density is induced in this way. When the system reaches
the critical time, the current density develops a delta-function
singularity, leading to a configuration with a polarized
fundamental string - anti-fundamental string pair near the
critical time.

\section{Effective Field Theory Approach\label{section4}}
It is also of interest to compare the results obtained from the
boundary-state approach with those from other approaches. In this
section, we analyze the D-brane decay in the effective field
theory approach, extending previous results \cite{reysugimoto} to
the case with inhomogeneous tachyon rolling. Consider the
world-volume Lagrangian of an unstable D$p$-brane:
\begin{eqnarray}
\cL_{\rm DBI} =-\cT_p \,V(T)\sqrt{-\det(\eta+F)}\,\cF(z),
\label{lag}
\end{eqnarray}
where $V(T)$ is a tachyon potential and $\cF(z)$ is a function of
\begin{eqnarray}
z:=\l({1 \over  \eta+F}\r)^{(ab)}\partial_a T\partial_b T,
\nonumber
\end{eqnarray}
normalized such that $\cF(0)=1$. The energy-momentum tensor
$T_{ab}$ and the Kalb-Ramond current tensor $\cQ_{ab}$ are
obtainable by replacing $\eta_{ab}\ra\eta_{ab}+h_{ab}$, $F_{ab}\ra
F_{ab}+b_{ab}$ in the Lagrangian Eq.(\ref{lag}) and expanding in
powers of $h_{ab}, b_{ab}$:
\begin{eqnarray}
\delta\cL_{\rm DBI} &=&\frac{\cT_p}{2}
\,V(T)\sqrt{-\det(\eta+F)}\,\,\cD^{ab} \,(h_{(ab)}+b_{[ab]})+\cdots\nn
&\equiv&\half\l( T^{(ab)}h_{(ab)}+\cQ^{[ab]}b_{[ab]}\r) + \cdots,
\label{TQ}
\end{eqnarray}
where
\begin{eqnarray}
\cD^{ab}=\cF(z)\l(\frac{-1}{\eta+F}\r)^{ab}+2\cF'(z)
\l(\frac{1}{\eta+F}\r)^{ac}\del_c T \del_d
T\l(\frac{1}{\eta+F}\r)^{db}.
\end{eqnarray}

Here, we consider rolling tachyon field with spatial modulation in
$x^1$-direction,
\begin{eqnarray}
\del_0 T\equiv\dot T,~~~\del_1 T\equiv T',~~~\del_2
T=\cdots=\del_p T=0, \label{D} \end{eqnarray}
and, for simplicity, turn off all other components of the gauge
fields but $F_{02}=e$. Then, Eq.(\ref{TQ}) and Eq.(\ref{D}) imply
that nonzero components of the energy-momentum tensor $T_{ab}$ are
\begin{eqnarray}
T_{00}&=&+\cT_p\,\gamma\, V(T)\l( \cF(z)+ 2\gamma^2 {\dot T}^2 \,
\cF'(z)\r),\nn T_{01}&=&+\cT_p\,\gamma\, V(T)\, 2 \dot T
T'\cF'(z),\nn T_{11}&=&-\cT_p\,\gamma^{-1}\,V(T)\l(
\cF(z)-2T'^2\cF'(z)\r),\nn T_{22}&=&-\cT_p\,e^2\gamma\,V(T)\l(
\cF(z)+ 2\gamma^2 {\dot T}^2 \, \cF'(z)\r)
-\cT_p\gamma^{-1}\,V(T)\cF(z),\nn
T_{ii}&=&-\cT_p\,\gamma^{-1}V(T)\cF(z), \qquad({\rm for}\quad
i=3,4,\dots,p), \label{effT}
\end{eqnarray}
and nonzero components of the Kalb-Ramond current tensor ${\cal
Q}_{[ab]}$ are
\begin{eqnarray}
\cQ_{02}&=&e\cT_p\,\gamma\, V(T)\l( \cF(z)+ 2\gamma^2 {\dot T}^2
\, \cF'(z)\r),\nn \cQ_{12}&=&e\cT_p\,\gamma\, V(T) \, 2 \dot T
T'\cF'(z), \label{effQ}
\end{eqnarray}
with $z=(-\gamma^2 {\dot T}^2 +{T'}^2)$ and
$\gamma=1/\sqrt{1-e^2}$.

In the effective field theory approach to unstable D-brane based
on the Lagrangian Eq.(\ref{lag}), second and higher derivatives of
the tachyon and gauge potential are neglected. As such, in case
the tachyon evolves sharply and rapidly, we would not expect a
good agreement with the corresponding results of the previous
section, where the boundary state approach surely indicated rapid
and sharp evolution of the tachyon field. Despite such
shortcomings anticipated, we see in the above result a behavior
observed in the boundary-state approach in Eq.(\ref{Te}) and
Eq.(\ref{Qe}). For example, relations
\begin{eqnarray}
\cQ_{02}=eT_{00}, \qquad \cQ_{12}=eT_{01}, \qquad T_{22}=-e^2
T_{00}+T_{ii}. \label{QeT}
\end{eqnarray}
featured by Eqs.(\ref{Te}, \ref{Qe}) also follow from
Eqs.(\ref{effT}, \ref{effQ}). In particular, we again obtain from
the first equation in Eq.(\ref{QeT}) that the energy density is
proportional to the fundamental string charge density.

One might think from the above observation Eq.(\ref{QeT}) that, in
so far as the electric field $e$ is kept fixed, any configuration
is classically stable since energetics does not tell which one is
favored among those with a fixed total fundamental string charge.
However, this is not correct. As we have seen explicitly in the
previous section, the system does evolve to thin electric flux
tubes without changing total energy and fundamental string charge
during the conversion process. The reason is as follows. In the
conventional stability analysis, one compares the energy of static
configurations with a fixed total charge. If one finds a minimum
energy configuration among them, this configuration is considered
stable. This is because the kinetic energy is always
positive-definite and the minimum energy configuration cannot
evolve dynamically to other ones without increasing the total
energy. Note that, in this argument, one has implicitly assumed
that variation of the fields does not affect the total charge. In
our situation, however, it is not enough to consider the energy of
static configurations with a fixed fundamental string charge, as
the kinetic energy \footnote{more generally, contribution of
time-dependence of the tachyon to the total energy} is also seen
to contribute to the fundamental string charge. Even if one finds
a configuration of a minimum energy among the static ones with a
fixed charge, it is always possible for the configuration to
evolve to some other configuration with ``less" fundamental string
charge, (here we mean ``less" provided we set $\dot T=0$ by hand),
as the rest of the fundamental string charge can be carried off by
the kinetic energy. This is to be contrasted against the reasoning
of \cite{yietal2,kawaikuroki}, where the conventional stability
analysis was taken for granted.

If we allow the electric field $e$ to vary, the minimum energy
configuration with a fixed total fundamental string charge is
realized in the extremal limit $|e|\ra 1$. Note that, in this
limit, one also needs to set $V(T)\ra 0$ from the outset to keep
the total energy finite \footnote{Here, we have assumed that the
kinetic energy is positive-definite, $i.e.$ $\cF(z)+2\gamma^2{\dot
T}^2 \cF'(z)\ge\cF(0)=1$.}. This is a special limit in which the
time evolution is frozen because of the (by now well-understood)
time-dilation effect of the electric field, and can be easily seen
from the fact that this system is T-dual to a codimension-one
unstable D-brane moving with the speed of light
\cite{reysugimoto}. In this limit, it is claimed that, since the
system does not evolve at all, one can engineer a string fluid of
an arbitrary energy density distribution \cite{yietal2}. On the
other hand, if one starts from a generic point of the potential
hill and consider the dynamical decay of the unstable D-brane,
there is no reason that the system would evolve to a string fluid
of arbitrary energy distribution advocated in \cite{yietal2}.
Indeed, we shall now show that the system evolves to the contrary.

Suppose we start with an inhomogeneous tachyon configuration with
$V(T)\ne 0$ and $|e|<1$, as considered above. The pressure along
the $x^1$-direction, $p_1\equiv T_{11}$, is, in general, non-zero
and gets modulated along the $x^1$-direction. Then, the momentum
density along the $x^1$-direction, $T_{01}$, cannot remain
constant, so the energy (and hence fundamental string charge
density) distribution would change every moment during the
evolution. In order to see whether the flux will be squeezed and
form an arbitrarily thin string-like flux tube or diffuse to a
uniform distribution, it is imperative to follow the dynamics of
the system carefully.

As an example, let us examine the background-independent string
field theory action for non-BPS D-branes in superstring theory. We
have \cite{KuMaMo}
\be V(T) = e^{ - {1 \over 2} T^2} \qquad {\rm and} \qquad \cF(z) =
{z 4^z \Gamma^2 (z) \over 2 \Gamma (2z)}. \label{BSFT} \ee
If we take a linear-gradient profile of the tachyon field $T=ux^1$
and $\dot T=0$ as an initial tachyon configuration,
the energy density $\rho$ and the pressure $p$ along the
$x^1$-direction are
\begin{eqnarray}
\rho&=&T_{00}=+ \cT_p\gamma\, V(ux^1)\cF(u^2), \nn
p_1&=&T_{11}=-\cT_p\gamma^{-1}\, V(ux^1)\cD(u^2), \nonumber
\end{eqnarray}
where $\cD(z):=\cF(z)-2z\cF'(z)$. In this case, we see that the
$x^1$-coordinate dependence is carried solely by the tachyon
potential, $V(u x^1)$. One can show that the function $\cF(u^2)$
is always positive for any value of $u^2>0$, so the energy density
$\rho$ is manifestly positive-definite. The energy density $\rho$
is accumulated around the origin, $x^1=0$, with a typical width of
order ${\cal O}(u^{-1})$. It was shown in \cite{KuMaMo} that the
total energy decreases monotonically with $u$. The energy density
becomes delta-function distribution $\rho(x^1)\propto \delta(x^1)$
in the $u\ra\infty$ limit, representing a BPS D$(p-1)$-brane with
the electric field. The function $\cD(u^2)$ is also a positive
function with $\lim_{u\ra\infty}\cD(u^2)=0$. Therefore, in so far
as $u^2$ is finite, the pressure $p_1$ is negative-definite with
the minimum located at the maximum of the energy density. From the
continuity equation $\del_0 T_{01}+\del_1 p_{1}=0$, we see that
the momentum density $T_{01}$ is turned on, and the energy starts
to accumulate to the center. This clearly suggest that the the
system will evolve to a thin string configuration. In order to
learn detailed time evolution of the system, one will need to
solve the equations of motion and follow the evolution explicitly.

In the above analysis, we considered the evolution under the
initial condition of a kink-like tachyon profile. It carries a
D$(p-1)$-brane charge, and hence the decay product would be a
bound state of the D$(p-1)$-brane and fundamental string
constituents. In order to obtain purely fundamental string
constituents in the final state, we will need to consider a
configuration with $T''\ne0$. In this case, it is difficult to
analyze the system based on the effective action Eq.(\ref{lag}),
as the second and higher derivatives of the tachyon field are
assumed small and thus truncated. Qualitatively, however, until
the tachyon modulation has not grown up large enough, the energy
density and the pressure of the system are well approximated by
Eq.(\ref{effT}). Recalling that the qualitative behavior is
governed primarily by the tachyon potential $V$, we expect that
the pressure $p$ reaches its minima at locations where the energy
density reaches its maxima. As such, again, we expect that the
inhomogeneity tend to grow as above. (See
\cite{starobinsky,Mukoh,cline,IsUe2} for related arguments.)

\section{Discussion}
In this paper, to gain deeper understanding concerning dynamical
fate of the fundamental string constituents, we have studied
rolling of the inhomogeneous tachyon in the presence of constant
gauge fields on the world-volume of the decaying D-brane. Quite
interestingly, we have found that the fundamental string charge
density is induced not only by the electric field but also by the
magnetic field. Although the gauge fields turned on are uniform,
distribution of the fundamental string constituents is spatially
inhomogeneous because of modulation of the rolling tachyon field.
We have shown that the tachyon rolling defines an exactly marginal
boundary perturbation to the boundary state and have solved the
system exactly. It represents an exact classical solution of
string theory, at least prior to reaching the critical time, where
the system develops a singularity.

The most pressing point for further study is a physical
interpretation of the critical time and singularity therein.
Notice that our consideration corresponds to the weak coupling
perturbation theory at the lowest order, and we ignored radiation
of massless and massive closed string modes, triggered by the
rolling tachyon, as well as back-reaction to the decaying D-brane
and background geometry.
Indeed, we have seen that the total energy is conserved and stored
in the open string sector until the critical time. The singularity
seems to suggest that the weak coupling limit is not a smooth
limit to take. Once we turn on $g_{\rm st}$, no matter how tiny it
is, the coupling to closed string modes blow up at the critical
time. So, the physics with $g_{\rm st} =0$ and that with tiny
$g_{\rm st}$ could be completely different at the critical time
\footnote{See \cite{okudasugimoto} for related issues for the late
time behavior of the homogeneous rolling tachyon.}.

In the more realistic set-up with small but non-zero $g_{\rm st}$,
inferring from Eq.(\ref{sfteq}), it is plausible to suppose that
the energy will be transferred to a collection of closed strings
near the critical time. After some energy is converted to closed
string modes, it would appear as if the energy is lost for the
observer looking only at the open string modes. It is not clear
whether the energy is radiated away or stored at the singularity
to form a macroscopic fundamental string or a D-brane. Once we
turn on non-zero $g_{\rm st}$, since the coupling to closed string
becomes impulsive and huge near the singularity, there ought to be
a large back reaction. In particular, it is not hard to imagine
that the space-time around the energy core will be severely curved
to form an event horizon. If it is the case, the energy cannot
escape through it, at least classically. This could be a possible
scenario for the formation of a macroscopic fundamental string or
a D-brane. If the string coupling $g_{\rm st}$ is nonzero (no
matter how small), nonlinear coupling of the tachyon to other open
string and closed string states becomes important near the
singularity. It would be certainly of interest to take these
string states into consideration, and obtain a more complete
picture of the rolling tachyon dynamics.

In \cite{yietal1}, it was claimed that the non-perturbative
confinement mechanism is responsible for the formation of
string-like object in unstable D-brane system. It was further
argued in \cite{yietal2,yietal3} that classical dynamics has no
mechanism to make the electric flux tube thin. In
\cite{kawaikuroki}, classical stability analysis for the unstable
D-brane effective action was made, and it was claimed that the
flux tube is unstable against uniform spread-out. However, our
result suggest that the flux can squeeze dynamically to a thin
flux tube {\sl even at the classical level}, in contrast to the
conclusion drawn in \cite{yietal2,yietal3,kawaikuroki}. Note that
 \cite{yietal2,yietal3} considered the situation that the
tachyon potential is set to zero from the outset (instead of
scaling limit), so it is delicate to make a direct comparison of
their results with ours in which we considered the tachyon rolling
down from a generic point of the potential hill.

Furthermore, in the works \cite{yietal2, kawaikuroki},
(in)stability of the flux tube was analyzed by looking for a
static configuration minimizing the total energy among those with
a fixed fundamental string charge. With the identification of the
correct electric displacement field, one readily sees that all
configurations with the same fixed charge have the same total
energy, so it appears superficially that {\sl any} of these
configurations is classically stable. However, as we have argued
thoroughly in section \ref{section4}, this stability criterion
actually does not tell us anything about the stability of the flux
distribution in our case. This is because, for the decaying
D-brane, it should be noted that the kinetic energy (in addition
to the gradient energy) of the tachyon field carries the
fundamental string charge as well. Consequently, in general, the
system does evolve even for the total charge held fixed, and this
is the physical reason behind the conclusion drawn in this work
differently from \cite{yietal2, yietal3, kawaikuroki}.

How generic is the conclusion we have drawn? We have analyzed in
detail the tachyon rolling of the special configuration given in
Eq.(\ref{c2bdryint}), and observed the formation of the
string-like thin electric flux tube. We also pointed out that
modulation of the tachyon field tends to grow by using both the
boundary-state and the effective field theory descriptions. On the
other hand, not every inhomogeneous tachyon configuration will
eventually evolve to a configuration with thin flux tubes like
this. In fact, we can easily construct an example with static
inhomogeneous energy density by turning on the tachyon profile of
the form:
\begin{eqnarray}
T(x^0,x^1)=\lambda \cosh(x^0)+\lambda' \cos(x^1). \nonumber
\end{eqnarray}
This profile also induces an exactly solvable boundary
perturbation. Thus, in order to see the fate of unstable D-branes
with generic tachyon configuration, a more general analysis is
desirable, which we leave for future study.

Finally, \cite{hoetal} has proposed an another approach in which
the dynamical formation of fundamental string in unstable D-brane
is argued within the context of classical effective field theory.
It would also be interesting to elaborate relation of theirs to our
approach.

\section*{Acknowledgement}
We thank J. Ambjorn, K. Hashimoto, R.A. Janik, H. Kawai, O. Lunin,
J. Maldacena, F. Sugino, and S. Terashima for discussions. This
work was carried out while SJR was a member of the Institute for
Advanced Study. SJR thanks the School of Natural Sciences for the
hospitality and the Fellowship. SS is supported in part by Danish
Natural Science Research Council.

\end{document}